\documentclass[journal,12pt,draftclsnofoot,onecolumn]{IEEEtran}
\usepackage{amsmath,amssymb,amsfonts}
\usepackage{algorithmic}
\usepackage{algorithm}
\usepackage{array}
\usepackage[caption=false,font=normalsize,labelfont=sf,textfont=sf]{subfig}
\usepackage{textcomp}
\usepackage{stfloats}
\usepackage{url}
\usepackage{verbatim}
\usepackage{color}
\usepackage{graphicx}
\usepackage{cite}
\begin{document}

\title{\LARGE Deep Reinforcement Learning Based Intelligent Reflecting Surface Optimization for TDD Multi-User MIMO Systems }

\author{Fengyu Zhao, Wen Chen\thanks{(\it Corresponding author: Wen Chen.)}, Ziwei Liu, Jun Li, and Qingqing Wu

\thanks{Fengyu Zhao, Wen Chen, Ziwei Liu, and Qingqing Wu are with the Department of Electronic Engineering, Shanghai Jiao Tong University, Shanghai 200240, China (e-mail: Aaronzfy@sjtu.edu.cn; wenchen@sjtu.edu.cn; ziweiliu@sjtu.edu.cn; qingqingwu@sjtu.edu.cn).}
\thanks{Jun Li is with the School of Electronic and Optical Engineering,
Nanjing University of Science Technology, Nanjing 210094, China (e-mail:
jun.li@njust.edu.cn).}}



\maketitle

\begin{abstract}
 In this letter, we investigate the discrete phase shift design of the intelligent reflecting surface (IRS) in a time-division duplexing (TDD) multi-user multiple-input-multiple-output (MIMO) system. We modify the design of deep reinforcement learning (DRL) scheme so that we can maximizing the average downlink data transmission rate free from the sub-channel channel state information (CSI). Based on the characteristics of the model, we modify the ``proximal policy optimization (PPO)" algorithm and integrate gated recurrent unit (GRU) to tackle the non-convex optimization problem. Simulation results show that the performance of the proposed PPO-GRU surpasses the benchmarks in terms of performance, convergence speed, and training stability.
\end{abstract}


\section{Introduction}
\IEEEPARstart{I}{ntelligent} reflecting surface (IRS) is a low power technology that smartly tunes the radio signal prorogation in wireless networks via a plurality of low-cost passive reflecting elements. Numerous influential works have been done on the configuration of continuous phase shifts of IRS with different design objectives. The authors in \cite{b1} studied an IRS-aided radar-communication (Radcom) scenario considering the cross-correlation design and the interference introduced by the IRS on the Radcom base station (BS).
In \cite{b2}, IRS-assisted simultaneous wireless information and power transfer (SWIPT) non-orthogonal multiple access (NOMA) networks are investigated to minimize BS transmit power. In \cite{b3}, multiple access schemes are investigated in IRS-aided wireless-powered mobile edge computing (WP-MEC). However, all the aforementioned papers are based on the instantaneous/perfect channel state information (CSI) assumption. It is a practically difficult task to acquire the CSI of the channel between the IRS and its serving BS/users since IRS is passive. Secondly, previous IRS studies concentrate on continuous phase shifts at reflecting components, which are difficult to realize practically due to hardware limitations. Hence, we focus on discrete phase shifts of the IRS without reliance on sub-channel CSI.\par
Deep reinforcement learning (DRL) has been widely used to solve resource allocation problems in wireless networks. The optimization problems are transformed into the design of the Markov decision process (MDP). The major advantage of DRL is that the mobility of the wireless network (e.g., time varying channel, terminal mobility, and real-time control of IRS reflective elements) can be resolved during the process when the agent keeps interacting with the wireless environment. There have been various attempts to operate the IRS based on machine learning.
In \cite{b5}, multi-agent RL algorithm is firstly employed in multiple IRSs-assisted multi-user (MU) systems.
In \cite{b6}, a model-free control of IRS based on received pilot is accomplished with a modified version of double deep Q-network (DDQN) called DRL with extremum seeking control (ESC). However, the proposed schemes only compare to other non-DRL algorithms under different values of Rician factor, which is less persuasive. Secondly, the DRL with ESC doesn't compare with the single DRL without ESC in the experiments, so the performance improvement of ESC is not verified. Thirdly, it is incomprehensible that the large action space doesn’t achieve better results than the small action space in the simulation. Inspired by those, we carry out discrete phase shift design based on DRL in a TDD MU-MIMO system free from the instantaneous sub-channel CSI.\par
The primary innovations of this letter can be summarized as follows:
\begin{itemize}
    \item We introduced a novel approach for achieving discrete control of IRS in TDD multi-user MIMO systems. The key benefit of our design is that the IRS deployment enhances communication quality without relying on the instantaneous or statistical CSI of sub-channels.
    \item Based on the characteristics of our model, we have made appropriate modifications to the PPO algorithm, resulting in the creation of a modified version named \text{PPO-GRU}. The new algorithm incorporates three significant modifications, including:
    \begin{enumerate}
        \item We integrated Gated Recurrent Unit (GRU), an improved type of Recurrent Neural Network (RNN), into the original PPO network structures of both the actor and critic. This modification allows the actor and critic to handle two types of state information, namely channel gains and angles, and deal with their correlation in the time domain within TDD systems.
        \item Normalization: dynamic mean and variance values are maintained during simulation for all encountered states or advantages. The current state or advantage is then normalized accordingly.
        \item We incorporated a strategy entropy term into the actor's loss function, ensuring the strategy's entropy remains as large as possible while optimizing the actor's loss.
    \end{enumerate}
\end{itemize}

\textit{Notations}: A column vector is represented as a boldface lowercase letter, and a matrix is defined with a boldface capital letter. $\odot$ represents the Hadamard product of a matrix, $(\cdot)^H$ is the conjugate transpose operation. For a set $A$, $|A|$ denotes the number of elements in the set. For a complex valued vector $x$, $|x|$ denotes $L_1$ norm. The operator $\operatorname{diag}(\cdot)$ represents the diagonal matrix of a vector. The random variable $x$ following the complex Gaussian distribution with zero-mean and unit variance is represented as $x \sim \mathcal{C N}(0,1)$.
\begin{figure}[htb]
    \centering
    \includegraphics[width=0.5\textwidth]{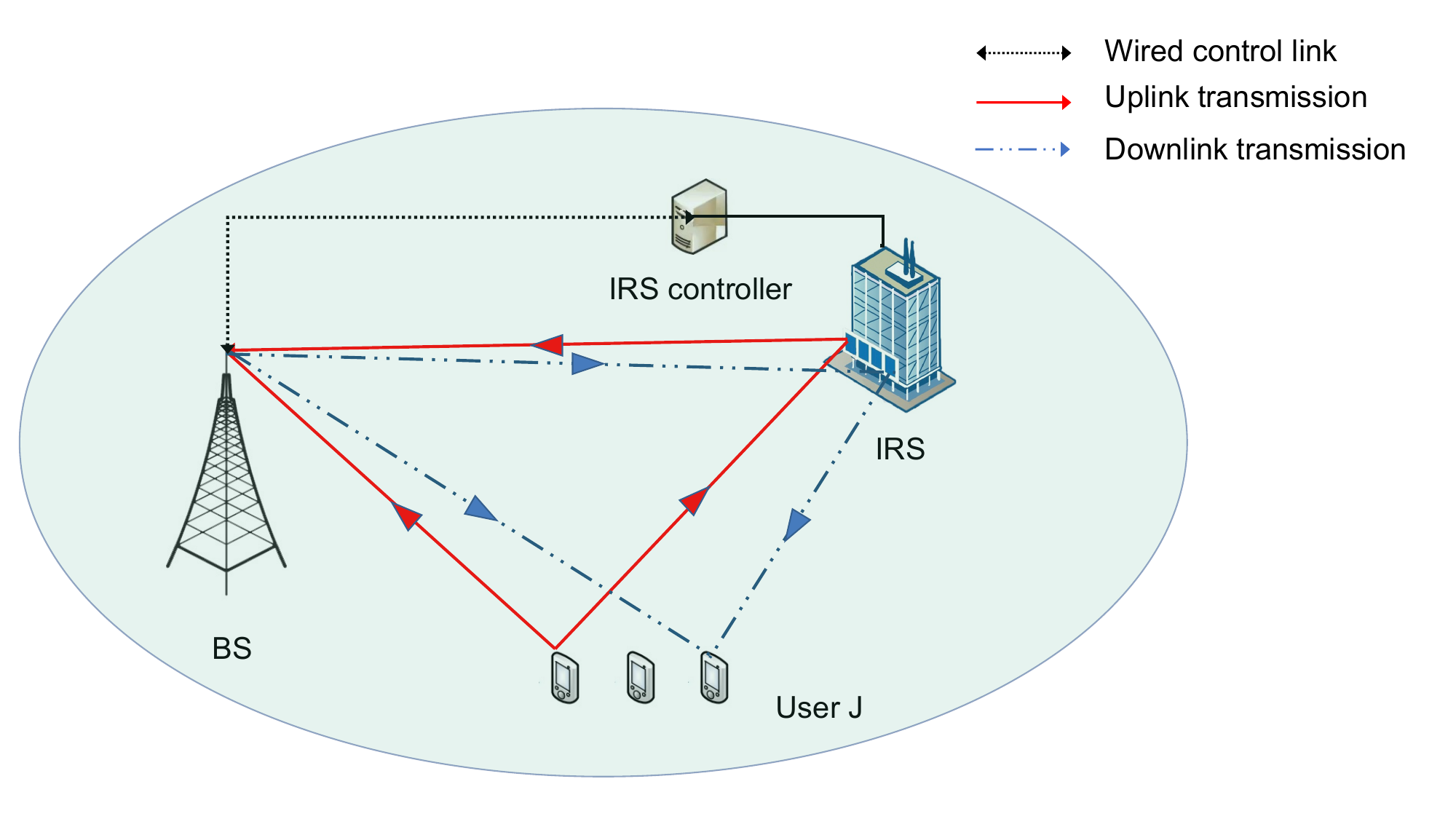}
    \caption{The TDD multi-user MIMO scenario.}
\end{figure}
\section{System Model and Problem Formulation}
 As shown in Fig. 1, the communication procedure can be divided into uplink channel estimation and downlink data transmission.

\begin{table}[b]
\caption{SUMMARY OF MAIN NOTATION}
\centering
\begin{tabular}{cc}
\hline Notation & Description \\
\hline$N_B$ & Number of BS antennas \\
$N_R$ & Number of IRS elements \\
$N_K$ & Number of users \\
$\mathbf{H}_{UB}$ & Channel gain between users and BS \\
$\mathbf{H}_{UR}$ & Channel gain between users and IRS \\
$\mathbf{H}_{RB}$ & Channel gain between IRS and BS \\
$\Phi$ & The reflection coefficients matrix of the IRS \\
$K$ & The Rician factor \\
$R$ & Downlink data sum rate \\
$\mathcal{F}$ & A set of discrete angle \\

\hline
\end{tabular}
\end{table}
\textcolor{blue}{\subsection{Channel Estimation}}
At uplink transmissions, multiple users transmit orthogonal pilot signals to BS simultaneously. The received signal matrix can be represented as 
\begin{equation}
\mathbf{Y} = \mathbf{X}\mathbf{H}+\mathbf{N},
\end{equation}
where $ \mathbf{X} \in \mathbb{C}^{N_K \times N_K}$ is the pilot pattern,  $\mathbf{N} \in \mathbb{C}^{N_K \times N_B}$ is the additive white Gaussian noise, which element follows the circularly symmetric
complex Gaussian (CSCG) distribution $\mathcal{C N}(0,1)$. \par
In an IRS-assisted wireless communication system, the uplink channel gain $ \mathbf{H}$ is given by
\begin{equation}
\mathbf{H} = \mathbf{H}_{UB}+\mathbf{H}_{UR}\mathbf{\Phi}\mathbf{H}_{RB},
\end{equation}
where $\mathbf{\Phi} =\operatorname{diag}\left\{\gamma_1 e^{j \beta_1} , \gamma_2 e^{j \beta_2}, \ldots, \gamma_{N_R} e^{j \beta_{N_R}}\right\}$ is the reflection coefficients matrix. $ \gamma_i$ and $\beta_i$ depict the amplitude and phase shift reflecting coefficient of IRS element $ i $ respectively. We consider the phase shift of each IRS reflecting element restricted to a finite number of discrete value $\mathcal{F}=\{0, \Delta \theta, \cdots, \Delta \theta(N_R-1)\}$, where $\Delta\theta = 2\pi / N_R $.\par 
Let $\mathbf{H}_{UB} \in \mathbb{C}^{N_K\times N_B}$, $\mathbf{H}_{UR} \in \mathbb{C}^{N_K\times N_R}$, $\mathbf{H}_{RB} \in \mathbb{C}^{N_R\times N_B}$  portray the channel gain from users to the BS,  the channel gain from users to the IRS, and the channel gain from the IRS to the BS respectively.
All the channels are modeled as the Rician channel \cite{b7}, we take $\mathbf{H}_{UB}$ as an example,
\begin{equation}\label{channel}
\mathbf{H}_{U B}=\sqrt{\frac{K}{K+1}} \mathbf{H}_{U B, L o S}+\sqrt{\frac{1}{K+1}} \mathbf{H}_{U B, N L o S},
\end{equation}
where $K$ is the Rician factor, $\mathbf{H}_{U B, L o S}$ denotes the deterministic line-of-sight (LoS) component, and $\mathbf{H}_{U B, N L o S} $ signifies the fading non-line-of-sight (NLoS) component.\par
To get rid of the dependence on sub-channel CSI, we use the minimum mean square error (MMSE) to estimate the channel, which can be formulated as \cite{b8} 
\begin{equation}
\hat{\mathbf{H}}=\mathbf{Y} \mathbf{X}^H\left(\mathbf{X} \mathbf{X}^H+\sigma_N^2 \mathbf{I}\right)^{-1}  .
\end{equation}    
\par
\textcolor{blue}{\subsection{Data Transmission}}
At downlink transmission, zero-forcing (ZF) precoding is performed according to the reciprocity between uplink and downlink channel. The precoding matrix $ \hat{\mathbf{A}}$ is represented as 
\begin{equation}
    \hat{\mathbf{A}} = [\hat{a_1}, \hat{a_2}, \ldots, \hat{a_K}],
\end{equation}
where $ \hat{a_k} $ is the \textit{k}th power normalized vector of
\newline
 $ (\hat{\mathbf{H}}^H\hat{\mathbf{H}})^{-1}\hat{\mathbf{H}}^{H}$.

The received signal at the \textit{k}th user can be expressed as
\begin{equation}
y_{k}=\mathbf{a}_k^H \mathbf{h}_k x_k+\sum_{j \neq k}^{N_K} \mathbf{a}_j^H \mathbf{h}_k x_j+n_k,
\end{equation}
where $x_k$ is the signal to be sent to the \textit{k}th user, and $n_k \sim \mathcal{C N}\left(0, \sigma_k^2\right)$ is the additive white Gaussian noise. Consequently, the signal-to-interference-plus-noise ratio (SINR) at the \textit{k}th user can be written as
\begin{equation}
S I N R_k=\frac{\left|\mathbf{a}_k^H \mathbf{h}_k\right|^2}{\sum_{j \neq k}^K\left|\mathbf{a}_j^H \mathbf{h}_k\right|^2+\sigma_k^2}.
\end{equation}\par
The achievable downlink data sum rate can be obtained as 
\begin{equation}\label{data_rate}
R=\sum_{k=1}^{N_K} \log _2\left(1+S I N R_k\right).
\end{equation}

We only consider the fully reflective IRS, so our optimization problem is formulated as 
\begin{equation}
    \begin{aligned}&\max _{\Phi(t)} \lim _{T \rightarrow \infty} \frac{1}{T} \sum_{t=0}^T R(t), \\ &\text { s.t. }~\gamma_i(t) = 1, \quad \forall i \in\left\{1,2, \ldots, N_R\right\}, \\  &~~~~~~\beta_i(t) \in \mathcal{F}, \quad \forall i \in\left\{1,2, \ldots, N_R\right\}.\end{aligned}
\end{equation}

\section{Deep Reinforcement Learning-based Solution}
\textcolor{blue}{After obtaining the uplink CSI of the current time slot, the IRS controller adjusts the current phase based on the CSI and the phase of the previous time slot to improve the downlink transmission rate.}
\subsection{MDP Formula}
\begin{enumerate}
    \item Environment: We treat the whole wireless communication system except for the IRS controller as the environment, and the working mechanism of the environment is incomprehensible to it.
    \item Agent: The IRS controller is served as the agent, which changes the configuration of IRS based on the performance feedback from the environment and the phase of each IRS element in the past.
    \item State: We define the state ${S_t}$ as the combination of the channel estimation and IRS phase at time slot $t-1$,

\begin{subequations}
    \begin{align} 
       {S_t} &= \{{S_t^{\Phi}}, {S_t^{H}}  \}, \\ {S_t^{\Phi}} & \triangleq \left[ \Re\{ {\Phi_{t-1}} \}, \Im\{{\Phi_{t-1}} \} \right], \\ 
       {S_t^{H}} & \triangleq \left[ \Re\{ {\hat{H}_{t-1}} \}, \Im\{{\hat{H}_{t-1}} \} \right].
    \end{align}
\end{subequations}
\item Action: The action is defined as the amount of change in phase  from ${\Phi_{t-1}} $,
\begin{equation}\label{phase}
    {\Phi_t} = {\Phi_{t-1}} \odot \Delta {\Phi_{t}}.
\end{equation}
The phase shift $\Delta {\Phi}$ is limited to the subset\ (or full set) of $N_R$ point discrete Fourier transform (DFT) vectors $v(k)$,
\begin{equation}
 v(k) = \left[ 1, e^{\frac{j\pi k}{N_R}},\ldots, e^{\frac{j\pi(N_R-1)k}{N_R}}  \right].
\end{equation} 
For an example, when the size of the action space $|A| = 2n + 1$, we could set action space  $A = \{ v(-n),v(-n+1),\ldots,v(0), v(1),\ldots, v(n) \}$. IRS Controller will select $v(k) \in A$ depending on its current policy function, then the phase shift at time slot $t$ will be
\begin{equation}
     \Delta {\Phi_t} = \operatorname{diag} \left \{ v(k) \right \}.
\end{equation}
Firstly, it has been demonstrated in \cite{b9} that utilizing an IRS with 2-bit phase shifters can achieve the same asymptotic squared power gain as the ideal scenario with continuous phase shifts. \textcolor{blue}{Therefore, discrete phase control based on DFT matrix is already sufficiently effective \cite{b10} \cite{b11}.} Secondly, increasing the number of IRS elements only increases the dimensionality of the $v(k)$ vector. The agent still selects from $A$ even when the size of the discrete phase set is fixed at $|A| = 2n + 1$. As a result, our design is capable of achieving good convergence performance even with a larger number of reflecting elements.
\item Reward: We adopt the downlink data sum rate in (\ref{data_rate}) as the reward.
\end{enumerate}
\subsection{IRS Control Using Improved PPO Algorithm}
Our modifications to PPO can be summarized as modifying the network architecture, normalizing state and advantage, and modifying the loss function equation. We will explain each of these modifications in the order listed. \par

Given that the state in our MDP contains two types of information, namely phase and channel gain, and is correlated in the time domain, we have made modifications to the actor and critic network structures in the traditional PPO algorithm to accommodate this model. The modified actor and critic network structures are similar in nature, we present the actor network as an example in Fig.2.

\begin{figure}[htb]
    \centering
    \includegraphics[width=0.3 \textwidth]{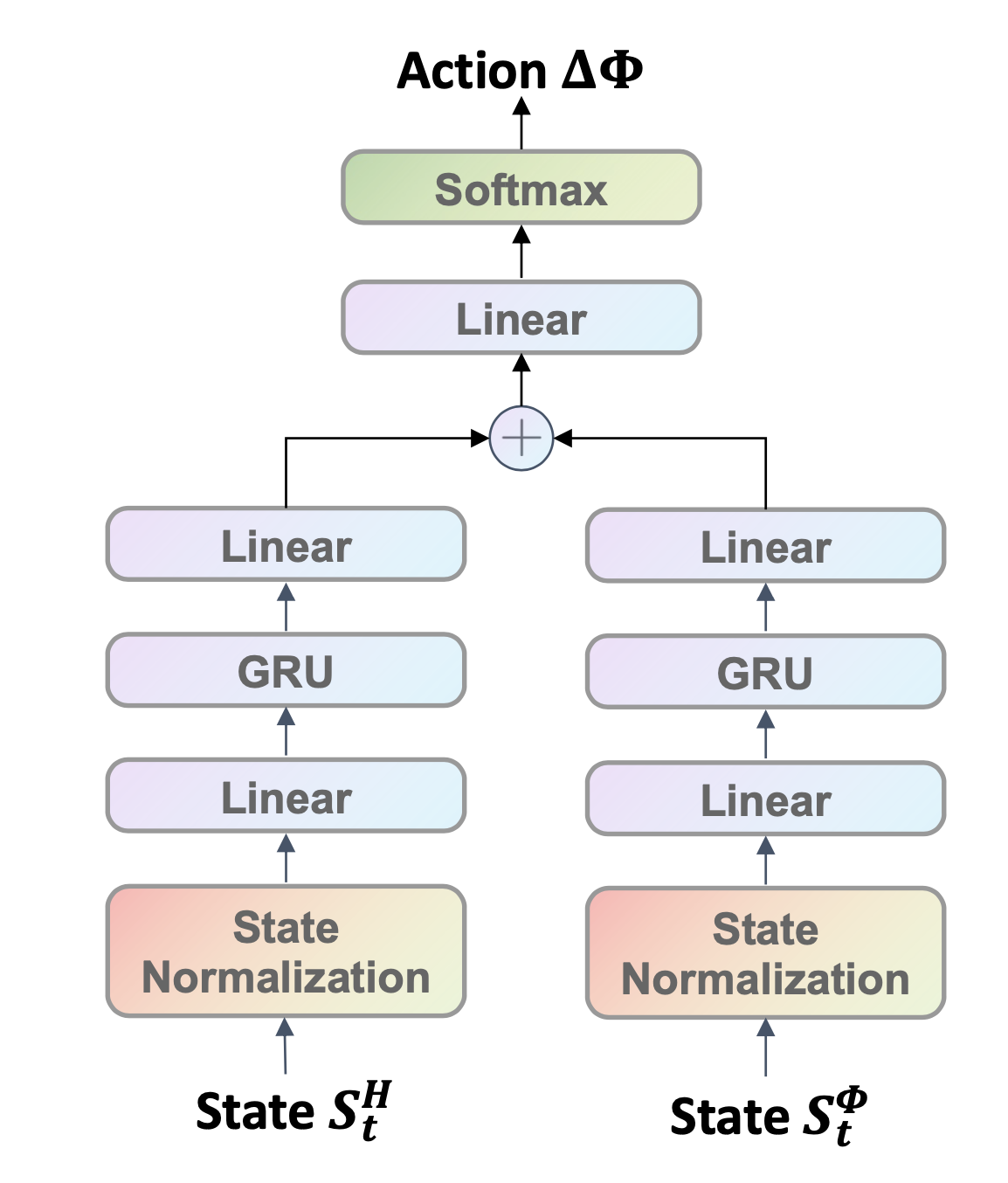}
    \caption{Revised PPO actor network framework.}
\end{figure}

Prior to inputting the states into the neural network, we first perform separate normalization for the two types of states. We demonstrate the normalization process using $S_t^{\Phi}$. At time slot $t$, all the values of the state ${S^{\Phi}} = [{S_1^{\Phi}}, {S_2^{\Phi}}, ..., {S_t^{\Phi}}]$ are recorded, and the normalized state is calculated using the following equation:

\begin{equation}\label{nor}
\hat{S_t^{\Phi}} = \frac{{S_t^{\Phi}} - \mu}{\sigma},
\end{equation}

where $\hat{S_t^{\Phi}}$ represents the normalized state of phase, $\mu$ is the mean of all the state values at time slot $t$, and $\sigma$ is the standard deviation.

In our system model, ${S_t^{\Phi}}$ and ${S_t^{H}}$ typically correlate with a prior data point or a data point spanning a time period. The LoS component within the channel gain experiences minimal variation if the user's position changes only slightly between time slots $t - 1$ and $t$, ensuring that the receiving antenna stays stationary. Furthermore, $\Phi_t$ is a time-dependent sequence data, as its value relies on the preceding phases ${\Phi_1}, {\Phi_2}, ..., {\Phi_{t-1}}$ as described in (\ref{phase}). Consequently, we incorporate two separate GRUs following two linear layers. The features extracted from the linear layer serve as inputs for the GRUs, enabling them to capture long-term dependencies in this sequence data. This enhances the model's accuracy and generalization capabilities. The results extracted from ${S_t^{\Phi}}$ and ${S_t^{H}}$ after passing through a three-layer network are added together and then input into another linear layer. In this way, the selected discrete phase of IRS is based on the consideration of the two types of state information.

Advantage is also normalized in our approach, which we refer to as mini batch normalization. After calculating the advantages using General Advantage Estimation (GAE) for a batch \cite{b9}, instead of directly normalizing the entire batch's advantages, we normalize the advantages of the current mini-batch before using it to update the policy in each iteration. Compared to the original PPO, our improved algorithm requires additional control over two hyperparameters: batch size $D$ and sample mini-batch $N_D$.
The loss function of actor $L_{actor}$ with mini batch advantage normalization is given by

\begin{equation}\label{a}
    L_{actor}=\min \left(\frac{\pi_\theta(a \mid s)}{\pi_{\theta_k}(a \mid s)} \hat{A}^{\pi_{\theta_k}}(s, a), g\left(\epsilon, \hat{A}^{\pi_{\theta_k}}(s, a)\right)\right) ,
\end{equation}
where 
\begin{equation}
    g(\epsilon, \hat{A}^{\pi_{\theta_k}}(s, a))= \begin{cases}(1+\epsilon) \hat{A}^{\pi_{\theta_k}}(s, a) & \hat{A}^{\pi_{\theta_k}}(s, a) \geq 0 \\ (1-\epsilon) \hat{A}^{\pi_{\theta_k}}(s, a) & \hat{A}^{\pi_{\theta_k}}(s, a)<0\end{cases},
\end{equation}
in which $\epsilon$ is a hyperparameter which roughly controls the variation between the new policy and the old one, and ${\hat{A}(s,a)}$ is the normalized advantage function. 

Thirdly, we modify the loss function expression above. Referring to the definition of entropy in information theory and probability statistics, the entropy of a strategy is represented as
\begin{equation}
    \mathcal{H}\left(\pi\left(\cdot \mid s_t\right)\right)=-\sum_{a_t} \pi\left(a_t \mid s_t\right) \log \left(\pi\left(a_t \mid s_t\right)\right).
\end{equation}
The greater the entropy of a strategy, the more evenly distributed the probabilities of selecting each action are. To improve the exploration capability of the algorithm, we add a term for strategy entropy to the actor's loss $L_{actor}$, multiply it by a coefficient $\delta$, and optimize $L_{actor}$ while maximizing the strategy's entropy. The modified loss function ${L_{actor}}'$ is given by
\begin{equation}\label{loss}
    {L_{actor}}' = L_{actor} + \delta * \mathcal{H}\left(\pi\left(\cdot \mid s_t\right)\right).
\end{equation}
where $\delta$ is the entropy coefficient.

\begin{algorithm}[htb]
	\renewcommand{\algorithmicrequire}{\textbf{Input:}}
	\renewcommand{\algorithmicensure}{\textbf{Output:}}
    \caption{Proximal Policy Optimization Based IRS Control in TDD Multi-User MIMO Systems.}
    \begin{algorithmic}
        \REQUIRE Initial IRS controller policy parameters $\theta_0$, initial value function parameters $\phi_0$. 
        \FOR{$k=0,1,2, \ldots$}
        \STATE Collect the trajectories into a set $\mathcal{D}_k=\left\{\tau_i\right\}$ by running current policy $\pi_k=\pi\left(\theta_k\right)$ in the environment.
        \STATE Randomly select a mini batch of trajectories. Compute rewards-to-go $\hat{R}_t = \sum_{t^{\prime}=t}^T R\left(s_{t^{\prime}}, a_{t^{\prime}}, s_{t^{\prime}+1}\right)$.
        \STATE Compute advantage estimates $\hat{A}_t$ using GAE method and do the mini batch normalization.
        \STATE Update the policy by minimizing the loss function defined in (\ref{loss}).

        \STATE Update value function by minimizing the mean-squared error between value function $V_\phi(s_t)$ and $\hat{R}_t$,
        $$
\phi_{k+1}=\arg \min _\phi \frac{1}{\left|\mathcal{D}_k\right| T} \sum_{\tau \in \mathcal{D}_k} \sum_{t=0}^T\left(V_\phi\left(s_t\right)-\hat{R}_t\right)^2.
$$
        \ENDFOR
        
    \end{algorithmic}
\end{algorithm}

\begin{table}[b]
\centering

\caption{Networks Parameters }
\begin{tabular}{|c|c|}
\hline Total Training steps $T$ & 10000000 \\
\hline Batch size $D$ & 2048 \\
\hline Sample mini-batch $N_D$ & 64 \\
\hline Discount factor $\gamma$ & 0.99 \\
\hline GAE parameter $\lambda$ & 0.95 \\
\hline Learning rate of actor network $\alpha_h$ & 0.0003 \\
\hline Learning rate of the critic network $\alpha_p$ & 0.0003 \\
\hline PPO clip parameter $\epsilon$ & 0.2 \\
\hline PPO entropy coefficient $\delta$ & 0.01 \\
\hline
\end{tabular}

\end{table}

\section{Simulation Results}
\subsection{Simulation Settings}
We establish our model in a three-dimensional Cartesian coordinate system. The BS is located at the coordinate [0,0,0], and IRS is placed at the coordinate [5,5,5]. The UEs whose height is ranging from 1.5m to 1.8m are uniformly distributed in a circle area with radium equal to 10m. The Rician factor $K $ = 10. The LoS component varies every 20 seconds by randomly selecting the user positions within the circle, while the NLoS component varies every second. \par
The channel matrix $\mathbf{H}_{UB},\mathbf{H}_{UR} $ and $\mathbf{H}_{RB}$ are generated in a similar manner, we take $\mathbf{H}_{RB}$ as an example. The LoS channel gain between IRS and BS is
\begin{equation}
\mathbf{H}_{R B, L o S}=\mathbf{v}_B \mathbf{v}_R^H,
\end{equation}

where the steering vectors are formulated as

$ \mathbf{v}_R=\mathbf{v}\left(\Psi_R, N_{R, y}\right)=\left[1, e^{j \pi \Psi_R}, \ldots, e^{j\left(N_{R, y}-1\right) \pi \Psi_R}\right]^T$, \\

$\mathbf{v}_B=\mathbf{v}\left(\Psi_B, N_{B, x}\right)=\left[1, e^{j \pi \Psi_B}, \ldots, e^{j\left(N_{B, x}-1\right) \pi \Psi_B}\right]^T. 
$
    
According to \cite{b12}, the directional cosines $\Psi_R, \Psi_B$ are represented as
\begin{subequations}
    \begin{align} 
       \Psi_R & =\mathbf{e}_{R}^T \mathbf{e}_{B R}, \\ \Psi_B & =\mathbf{e}_{B}^T \mathbf{e}_{B R}.
    \end{align}
\end{subequations}
We place the uniform linear array (ULA) of the BS at the coordinate [1,0,0], and assume that the reflector array of IRS can be regarded as a ULA placed at [0,1,0], so $e_R = [0,1,0]^T$  and $e_B = [1,0,0]^T$.
The NLoS components follow the complex Gaussian distribution $\mathcal{C N}(0,1)$. $\mathbf{H}_{UB}$ and $\mathbf{H}_{UR} $ are calculated in the same way.\par

\subsection{Comparisons With Benchmarks}
We set the number of users $N_K$ = 2, the number of BS antennas $N_B = 2$, the action space size of IRS controller $\lvert A \rvert$ = 5, and the number of the elements of IRS $N_R = 32$. Other network parameters settings are listed in Table II.\par

\begin{figure}[htb]
    \centering
    \includegraphics[width=0.45\textwidth]{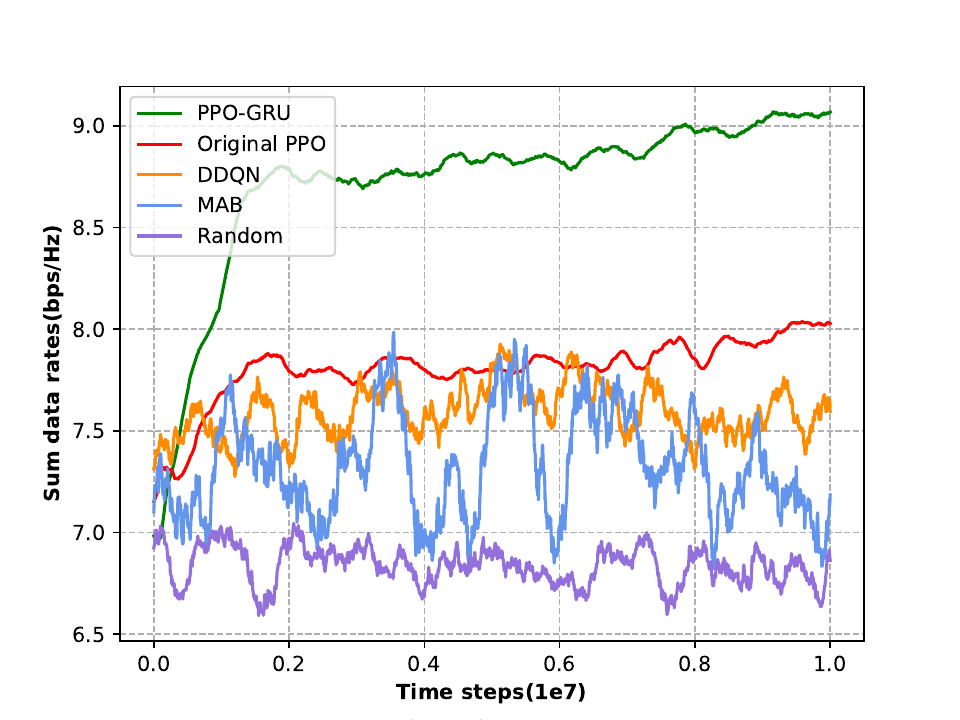}
    \caption{Performance comparisons of different algorithms.}
\end{figure}
In Figs. 3 and 4, we compare the proposed PPO-GRU scheme with four benchmarks: random reflection, multi-armed bandit (MAB), DDQN in \cite{b6} and original PPO. All the schemes use the same action sets but handle the information differently. Random reflection cannot analyze and utilize all kinds of information, resulting in the poorest performance. MAB is a simpler version of DQN which builds the connection between reward and action by calculating the reward distribution of all the arms. However, it fails to fully utilize the state information in our model. Compared to other DRL algorithms, such as the original PPO and DDQN, our PPO-GRU achieves better performance and faster convergence. Our modifications can enhance the model's accuracy and generalization capabilities, enabling the PPO-GRU to capture the changing dynamics of the wireless environment and adapt the agent's policy accordingly.

\begin{figure}[htb]
    \centering
    \includegraphics[width=0.45\textwidth]{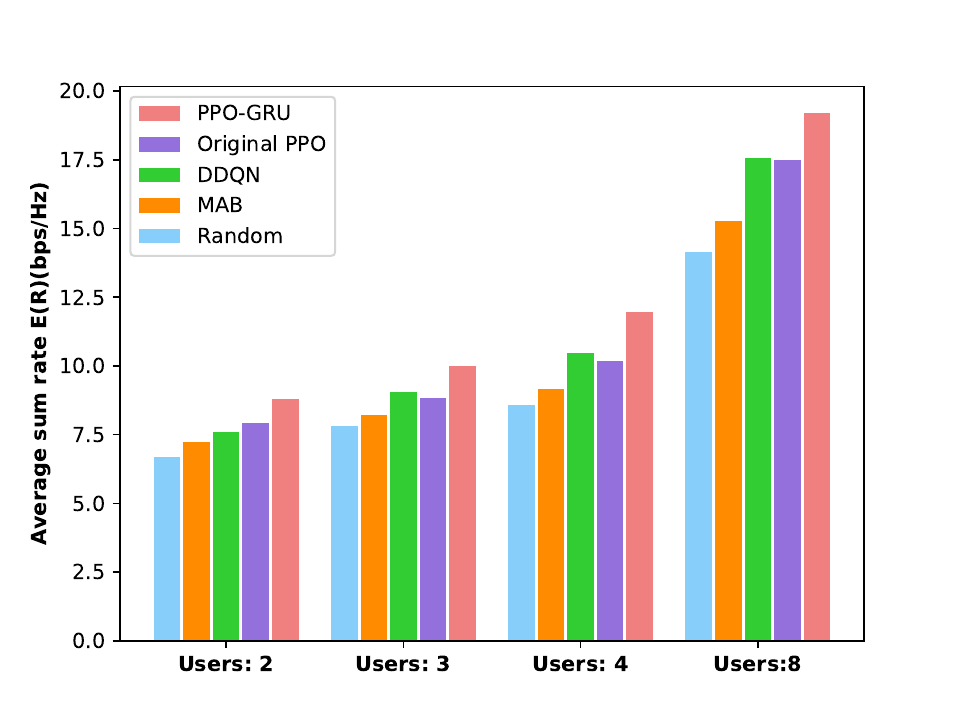}
    \caption{Comparisons of algorithm performance with different numbers of users}
\end{figure}

\begin{figure}[htb]
    \centering
    \includegraphics[width=0.45\textwidth]{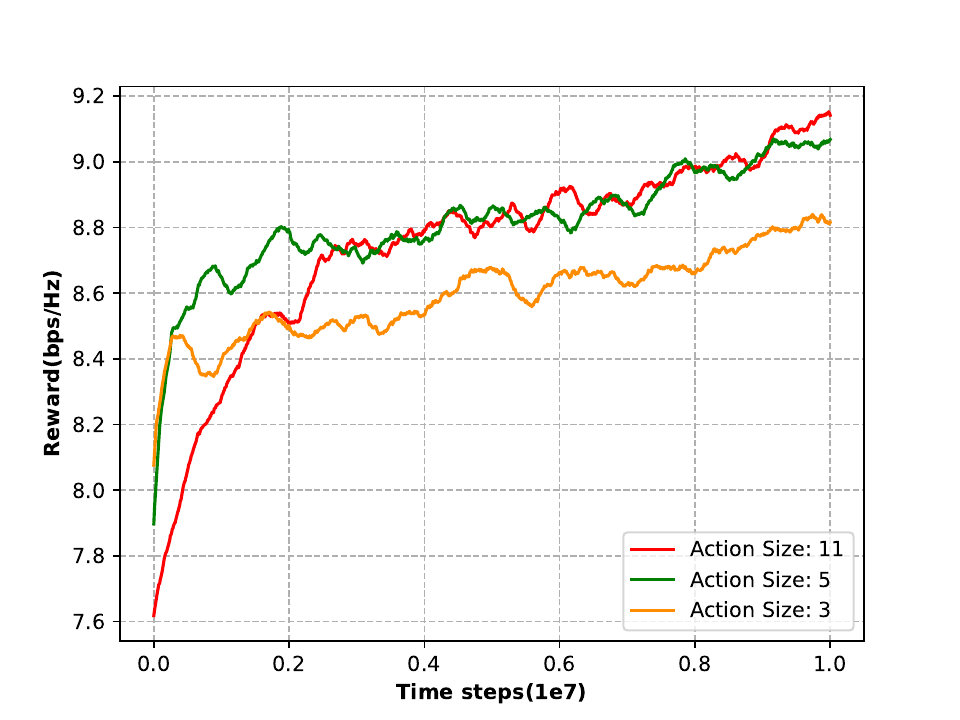}
    \caption{Performance of PPO with different action sets.}
\end{figure}

In Fig. 5, we study the impact of different action spaces $A$ on the performance of the proposed PPO scheme.  When $\lvert A \rvert$ = 3, the algorithm converges at the fastest speed but ends up with the lowest data rate. When $ \lvert A \rvert$ = 11, the agent performs poorly at the beginning of the training but results in the best performance.However, it will take too much time to converge if we increase the number of users.  When $ \lvert A \rvert$ = 5, it achieves nearly good performance and converges much faster. So a moderate size of action space is best for practical deployment.

\section{Conclusion}
This letter investigates the problem of maximizing the average data sum rate in IRS-assisted TDD MU-MIMO networks under the constraints of discrete phase shifts. To address this challenging problem, we propose an improved PPO algorithm. Simulation results demonstrate that our modified PPO algorithm outperforms previous algorithms in various scenarios. Moreover, we show that a well-designed action space can achieve both high training efficiency and good performance.


\begin{thebibliography}{00}
\bibliographystyle{IEEEtran}
\bibitem{b1}  M. Hua, Q. Wu, C. He, S. Ma, and W. Chen, ``Joint active and passive beamforming design for IRS-aided radar-communication," \emph{ IEEE Trans. Wireless Commun.}, vol. 22, no. 4, pp. 2278–2294, Apr. 2023.
\bibitem{b2} Z. Li, W. Chen, Q. Wu, K. Wang, and J. Li, ``Joint beamforming design and power splitting optimization in IRS-assisted SWIPT NOMA networks," \emph{IEEE Trans. Wireless Commun.}, vol. 21, no. 3, pp. 2019-2033, Mar. 2022.
\bibitem{b3} G. Chen, Q. Wu, W. Chen, D. W. K. Ng and L. Hanzo, ``IRS-aided wireless powered MEC systems: TDMA or NOMA for computation offloading?," \emph{IEEE Trans. Wireless Commun.}, vol. 22, no. 2, pp. 1201-1218, Feb. 2023.
\bibitem{b5} J. Zhang, J. Li, Y. Zhang, Q. Wu, X. Wu, F. Shu, S. Jin, and W. Chen, ``Collaborative intelligent reflecting surface networks with multi-agent reinforcement learning," \emph{IEEE J. Sel. Top. Signal Process.}, vol. 16, no. 3, pp. 532-545, Apr. 2022.
\bibitem{b6} W. Wang and W. Zhang, ``Intelligent reflecting surface configurations for smart radio using deep reinforcement learning," \emph{IEEE J. Sel. Areas Commun.}, vol. 40, no. 8, pp. 2335-2346, Aug. 2022.
\bibitem{b7} Andrea Goldsmith, Wireless Communications. Cambridge University Press, USA, 2005.
\bibitem{b8} H. Poor, An introduction to signal detection and estimation (2nd ed.). Springer-Verlag, Berlin, Heidelberg, 1994.
\bibitem{b9} Q. Wu and R. Zhang, ``Beamforming optimization for intelligent
reflecting surface with discrete phase shifts," in \emph{ Proc. IEEE Int. Conf.
Acoust. Speech Signal Process. (ICASSP)}, Brighton, UK, May 2019,
pp. 7830–7833.
\textcolor{blue}{\bibitem{b10} T. L. Jensen, and E. De Carvalho, ``An Optimal Channel Estimation
Scheme for Intelligent Reflecting Surfaces Based on a Minimum
Variance Unbiased Estimator," in \emph{ Proc. IEEE Int. Conf.
Acoust. Speech Signal Process. (ICASSP)}, pp. 5000-5004,
Barcelona, Spain, 2020.}
\textcolor{blue}{\bibitem{b11}Z. Sun and Y. Jing, ``On the performance of multi-antenna IRS-assisted
NOMA networks with continuous and discrete IRS phase shifting," in \emph{ IEEE
Trans. Wireless Commun.}, vol. 21, no. 5, pp. 3012–3023, May 2022.}
\bibitem{b12} J. Schulman, P. Moritz, S. Levine, M. Jordan, and P. Abbeel,  ``Highdimensional continuous control using generalized advantage estimation," in \emph{ Proc. Int. Conf. Learn. Representations,} 2016.
\bibitem{b13} W. Wang and W. Zhang, ``Jittering effects analysis and beam training design for UAV millimeter wave communications," \emph{IEEE Trans. Wireless Commun.}, vol. 21, no. 5, pp. 3131-3146, May. 2022.

\end{thebibliography}
\end{document}